# Comparison between the HUBCAP and DIGITBrain Platforms for Model-Based Design and Evaluation of Digital Twins[*]

Prasad Talasila[2][0000-0002-8973-2640], Daniel-Cristian Crăciunean[1][0000-0002-6725-3796],
Pirvu Bogdan-Constantin[1][0000-0003-3961-4539],
Peter Gorm Larsen[2][0000-0002-4589-1500],
Constantin Zamfirescu[1][0000-0003-0128-2436], Alea Scovill[3][0000-0002-3534-8758]

[1] Lucian Blaga University of Sibiu, Romania
{daniel.craciunean,bogdan.pirvu,constantin.zamfirescu}@ulbsibiu.ro
[2] DIGIT, Aarhus University, Aarhus, Denmark
{prasad.talasila,pgl}@ece.au.dk
[3] Agrointelli ApS, Aarhus, Denmark
als@agrointelli.com

**Abstract.** Digital twin technology is an essential approach to managing the lifecycle of industrial products. Among the many approaches used to manage digital twins, co-simulation has proven to be a reliable one. There have been multiple attempts to create collaborative and sustainable platforms for management of digital twins. This paper compares two such platforms, namely the HUBCAP and the DIGITbrain. Both these platforms have been and continue to be used among a stable group of researchers and industrial product manufacturers of digital twin technologies. This comparison of the HUBCAP and the DIGITbrain platforms is illustrated with an example use case of industrial factory to be used for manufacturing of agricultural robots.

**Keywords:** Digital twins . Industrial products . Model-Based Design . Co-simulation . HUBCAP . DIGITbrain . Agricultural robots . CPS.

## 1 Introduction

In order to optimize the use of industrial products such as manufacturing equipment, it is paramount to consider digital technologies. Here digital models, digital shadows and digital twins play an important role for the Cyber-Physical Systems (CPSs) and the manufacturing processes. Unfortunately, many of the software tools have rather high costs for getting started. This means that there is a risk that many Small and Medium Enterprises (SMEs) do not have enough financial power to join the digital transition. There is also a need for close collaboration between industrial product manufacturers, product users and CPS software providers so that reuse of CPS models and tools is possible. Thus, it is essential to establish sustainable and affordable collaboration platforms that lower the barriers to the use of digital twin technologies among SMEs.

---

[*] The work presented here is partially supported by the HUBCAP and the DIGITbrain projects. The HUBCAP and the DIGITbrain projects are funded by the European Commission's Horizon 2020 Programme under Grant Agreements 872698 and 952071 respectively.

In this paper we focus on and compare the platforms targeted by two H2020 Innovation Actions (IA) called HUBCAP [1,2,3] and DIGITbrain [4]. The focus of the HUBCAP platform is on Model-Based Design (MBD) for CPSs for any application domain. The HUBCAP collaboration platform contains interesting innovation called a sandbox enabling users to try MBD tools and models directly from an internet browser without having to install anything locally. The focus of the DIGITbrain platform is on evolving Digital Twins (DTs) for Manufacturing as a Service in a platform essentially enabling DTs from cradle to grave for industrial products. The main innovation here is a clear distinction between data, models and algorithms of a DT and in addition a high level of configurability on the execution environment.

We are going to take a closer look at the capabilities and differences of the HUBCAP and the DIGITBrain platforms. A case study of digital twin for industrial factory of agricultural robots has been used to illustrate the importance of collaborative MBD, publication, reuse, and evaluation of digital twins using the HUBCAP and the DIGITbrain platforms.

An important point to note is that the paper contains a review of the conceptual abstractions and implementations of two ongoing projects. Thus, the description provided in this paper is limited to capabilities implemented while writing this paper. The rest of this paper is organized as follows: Section 2 presents an overview of the HUBCAP and the DIGITBrain platforms, Section 3 compares these two digital twin platforms, and Section 4 deals with the Agrointelli's factory simulation case study.

## 2      The HUBCAP and the DIGITbrain Platforms

The HUBCAP and the DIGITbrain platforms are aimed at sustaining the ecosystem of CPSs for industrial users. The general approach taken is to help the users create, exchange and reuse DTs for industrial products (IPs). Both the projects have a marketplace for publishing, reuse of DT among manufacturers, consultants, software vendors and users.

Both the projects treat DT models and software as assets on the platform that can be reused. In the case of HUBCAP, the emphasis is on creation of models in general using the available software authoring tools (such models can for example be used in a DT context). The collaborative development of DT models is the primary advantage of the HUBCAP platform [2,3]. The HUBCAP platform contains HUBCAP Sandbox Middleware (HSM) which is a formally proven, secure platform for management of collaboration among users based on Sandbox concept [5,6].

The DIGITbrain platform enables creation of DTs for IPs using reusable models, algorithms / tools and data sources. The DIGITbrain platform, which is under development can host reusable data (D), models (M), algorithms (A), and Model-Algorithm (MA) pair assets [7,8]. The DTs are composed from these reusable assets and are evaluated on distributed computing infrastructure.

## 3   Comparison of the HUBCAP and the DIGITbrain Platforms

In this section, we provide a comparison of the workflows involved in creation and execution of a digital twin on both the HUBCAP and DIGITbrain platforms.

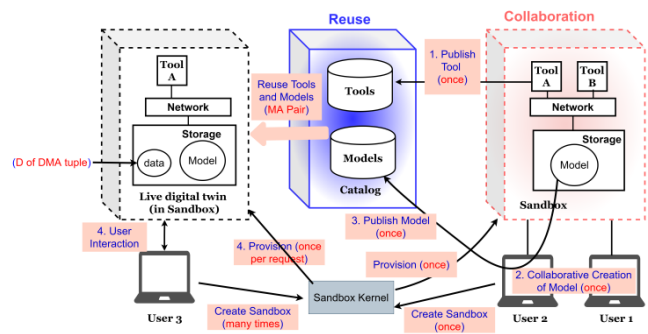

Fig. 1. Workflow in the HUBCAP project leading to execution of reusable digital twins.

Figure 1 shows a typical workflow in the HUBCAP platform if it is used in a DT context. The HSM is responsible for managing the workflow in the creation and execution of reusable DTs. The HSM manages collaboration among platform users in secure sandbox environments. The sandbox kernel, a part of HSM, is responsible for on-demand provisioning of a sandbox for users. Software providers can use a live sandbox to package their tool / software and publish the same into the catalog. The owner of a tool can also update and version their software. This is *step-1* in the workflow.

Each sandbox has a shared storage using which files can be exchanged in a sandbox. These usually are model, data, or log files that may be needed for collaboration within a sandbox. In *step-2*, user(s) typically develop a model, download the same to their computers before publishing it to the catalog (*step-3*). Model owners can also upload new versions of their models. In the HUBCAP project, the collaborative development (i.e., authoring) of tools and models is very easy. The publishing of a model or a tool need to be only once.

In *step-4*, users of a digital twin model can request the sandbox kernel to provision a sandbox in which model(s) and tool(s) needed for creation of a DT are available. Users can then upload required data to use the selected digital twin. Thus, even the HSM supports Data-Model-Algorithm (DMA) abstraction of a DT. Each sandbox has a unified web interface via which participants of a sandbox can collaborate. Thus, real-time collaboration between users is possible in both authoring and usage phases of digital twins. The provision of all sandboxes happens on a single server thus limiting the number of concurrent sandboxes that can be run.

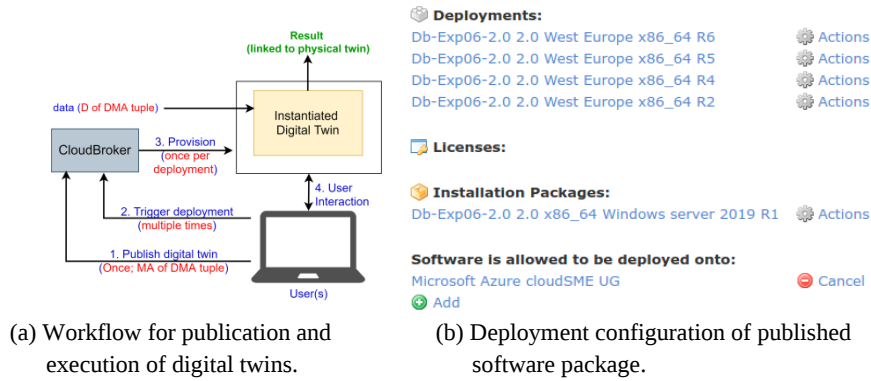

(a) Workflow for publication and execution of digital twins.

(b) Deployment configuration of published software package.

Fig. 2. The use of DIGITbrain project for evaluation of digital twins.

Figure 2(a) shows a typical workflow involved in the DIGITbrain platform (which itself is based on the CloudBroker platform [9]). The DIGITbrain platform allows for publishing of tool / software / algorithm package. The *first step* is the publication of model-algorithm pair on the CloudBroker platform. At present, the authoring of models and tools is outside the main part of the DIGITbrain platform.

One advantage of the DIGITbrain platform is the ability of users to use the published software package and configure the on-demand deployment of a DT. Fig. 2(b) shows different levels of configuring the deployment of a software package. The installation packages allow for the selection of the base computing environment (operating system, library dependencies, graphics hardware etc) and installation of the software package on the deployment environment. The deployment environment specifies the hardware computing infrastructure onto which the software is configured to be deployed. In *step-2*, users can trigger automated deployment of selected software via a web interface.

The selected deployment environment is usually available among competing cloud service providers. In the present case, the on-demand deployment happens on the Microsoft Azure - a public cloud service provider. It is also possible to select different kinds of hardware configurations and save them as deployment configurations. The DT users can select one of the possible deployments. In *step-3*, the CloudBroker platform provisions the software on selected computing infrastructure and makes the DT ready to use. Users can then import models, data or both into the live DT.

Table 1 summarizes the relative advantages of each implementation platforms explained above. An interesting side effect of both the implementations is a ready to use a DT published on the platforms. Users have the option to just publish the tools, or tools-models or tools-models-data. Publication of model, data within a packaged tool available on the platform might help users check the setup of tool with an example model-data pair before using their own data-model-algorithm tuple.

Table 1. Comparison on the relative advantages of implementing digital twins on the HUBCAP and the DIGITbrain platforms.

| Workflow | HUBCAP | DIGITbrain |
|---|---|---|
| Create new models | Yes | No |
| Collaboration between users | Yes | No |
| Publish and reuse models | Yes | No |
| Publish tools / software | Yes | Yes |
| Ease of publishing models and tools | Easy | Involves multiple steps each of which require significant domain knowledge |
| Reuse of models by customers | Yes | Yes |
| Deployment of software and model for instantiation of digital twins | Manual | Automated |
| Specification of deployment infrastructure | Not possible | Configurable in the publication phase |
| Execution environment | Single server | On scalable cloud resources |
| Export results to user | Manual | Supports both manual and automated |
| Reuse of evaluation environment | Yes, but manual deployment in each scenario | Nearly automated, click of a button |

## 4 Agrointelli Case Study

Agrointelli (AGI) is a Danish SME which produces customizable autonomous field robots that can perform various operations in a farmer's field (e.g., light tillage, seeding, weeding, spraying, etc.). As AGI increases their sales, it is expected that there will be spare part orders from existing customers which will cause interruptions in the production process. For this, AGI needs to optimize their production process while considering spare part orders from farmers and/or dealers. Thus, creating a production model that can be scaled up, in terms of detail and potential integration of real-time data (e.g., resource, process etc.) for realistic decision-making is critical for its future growth of AGI.

In this case study, a co-simulation model of factory production process has been created using the CPPS-SimGen tool. The INTO-CPS toolchain has been used as the tool for executing the factory co-simulation model [11]. Thus, the factory DT is used to simulate the behaviour of the manufacturing process for a period of several months. The factory manufacturing process is co-simulated at maximum production capacity. The result of the co-simulation is taken over by a Decision Support System (DSS) which outputs new production schedule. The decision process to update production schedule begins with the arrival of a new order for a set of broken parts. This new order will have to be included in the current production plan and completed as soon as possible without disturbing the ongoing production plan. By simulating the manufacturing process and using a DSS the estimated delivery date for the part orders is generated.

The factory co-simulation model has been created using the CPPS-SimGen tool running in a HUBCAP sandbox [10]. The factory model is then downloaded onto the user's computer. The model is published onto the HUBCAP models repository. This factory co-simulation model can be executed using the INTO-CPS toolchain, in a new sandbox on the HUBCAP platform. The required parts order data need to be uploaded onto a running sandbox. A new instance of factory DT can be created by using factory model and the INTO-CPS tool available in the HUBCAP catalogue.

In case of the DIGITbrain project, authoring of the factory model is not possible. In the DIGITbrain parlance, the factory co-simulation model is the DIGITbrain model and the INTO-CPS toolchain is the DIGITbrain algorithm. This model-algorithm (MA) pair is then published as a software package onto the CloudBroker platform. The published software package is configured with four deployment configurations and coupled with one commercial cloud service provider for on-demand execution of the factory DT. The required parts order data is directly uploaded into the cloud computer on which the factory DT is being executed. In this way, the Agrointelli use case has been used to demonstrate the capabilities of the HUBCAP and the DIGITbrain platforms.

## 5    Concluding Remarks

There is a timely need for collaborative and sustainable development platforms for DT stakeholders. The HUBCAP and DIGITbrain project platform are two prototype demonstrations of DT platforms. The HUBCAP platform is geared towards collaborative development and publication of models and tools that can be used for DTs. The DIGITbrain platform enables industrial product manufacturers, users, software providers and domain experts to collaborate on the publication and use of DTs. The comparative advantages of these two platforms have been demonstrated using an industrial case study. Even though there are overlaps in some of the features of these DT platforms, both are complementary in nature. Thus, the HUBCAP and the DIGITbrain platforms are important collaborative efforts to sustain the adoption of DT technologies among SMEs.

**Acknowledgements.** We would also like to express our thanks to the anonymous reviewers.

## References

1. V. Chapurlat and B. Nastov, "Deploying MBSE in SME context: revisiting and equipping Digital Mock-Up," in 2020 IEEE International Symposium on Systems Engineering (ISSE), 2020, pp. 1–8. doi: 10.1109/ISSE49799.2020.9272230.
2. Larsen,P.G.,Macedo,H.D.,Fitzgerald,J.,Pfeifer,H.,Benedikt,M.,Tonetta,S.,Marguglio, A., Gusmeroli, S., Jr., G.S.: A Cloud-based Collaboration Platform for Model-based Design of Cyber-Physical Systems. pp. 263–270. INSTICC, Proceedings of the 10th International Conference on Simulation and Modeling Methodologies, Technologies and Applications Volume 1: SIMULTECH (July 2020). https://doi.org/10.5220/0009892802630270
3. HUBCAP Project, https://www.hubcap.eu/, last accessed 2021/10/07.
4. DIGITbrain Project, https://digitbrain.eu/, last accessed 2021/10/07.
5. Kulik, T., Talasila, P., Greco, P., Veneziano, G., Marguglio, A., Sutton, L. F., ... & Macedo, H. D. (2021). Extending the Formal Security Analysis of the HUBCAP sandbox. Hugo Daniel Macedo, Casper Thule, and Ken Pierce (Editors), 36, Proceedings of the 19th International Overture Workshop, 1 October 2021, https://arxiv.org/abs/2110.09371.
6. Kulik, T., Macedo, H. D., Talasila, P., & Larsen, P. G. (2021). Modelling the HUBCAP Sandbox Architecture In VDM: A Study In Security. John Fitzgerald, Tomohiro Oda, and Hugo Daniel Macedo (Editors), 20, Proceedings of the 18th International Overture Workshop, 7 December 2020, https://arxiv.org/pdf/2101.07261.pdf.
7. Sandberg, M, et. al, Deliverable 5.1 Models for Digital Twins, DIGITBrain Project [internal communication], 30-June-2021.
8. Pena S., et. al, Deliverable 7.1 Minimum Viable Digital Product Brain, DIGITBrain Project [internal communication], 30-June-2021.
9. Kiss, T. "A Cloud/HPC Platform and Marketplace for Manufacturing SMEs." *11th International Workshop on Science Gateways, IWSG 2019*. 2019.
10. Functional Mock-up Interface for Model Exchange and Co-Simulation, Document version: 2.0.1 October 2nd 2019, https://fmi-standard.org/, accessed on 2021/10/21.
11. The INTO-CPS Tool Chain User Manual, Version: 1.3 Date: September, 2019, https://into-cps-association.readthedocs.io/en/latest/, accessed 2021/11/28.